\documentclass[aps,prstab,preprint,groupedaddress]{revtex4}
\usepackage{graphicx}
\newcommand{\be}{\begin{equation}}
\newcommand{\ee}{\end{equation}}
\newcommand{\bea}{\begin{eqnarray}}
\newcommand{\eea}{\end{eqnarray}}

\renewcommand{\vec}[1]{\textnormal{\boldmath$#1$}}

\begin{document}

\preprint{SLAC-PUB-9503}
%\preprint{arXiv/physics/0206003}
% Use the \preprint command to place your local institutional report
% number in the upper righthand corner of the title page in preprint mode.
% Multiple \preprint commands are allowed.
% Use the 'preprintnumbers' class option to override journal defaults
% to display numbers if necessary
%\preprint{}

%Title of paper
\title{Impedance of a Rectangular Beam Tube with Small Corrugations}

% repeat the \author .. \affiliation  etc. as needed
% \email, \thanks, \homepage, \altaffiliation all apply to the current
% author. Explanatory text should go in the []'s, actual e-mail
% address or url should go in the {}'s for \email and \homepage.
% Please use the appropriate macro for each each type of information

% \affiliation command applies to all authors since the last
% \affiliation command. The \affiliation command should follow the
% other information
% \affiliation can be followed by \email, \homepage, \thanks as well.
\author{K.L.F. Bane and G. Stupakov}
%\email[]{kbane@slac.stanford.edu}
%\homepage[]{Your web page}
\thanks{Work supported by the Department
of Energy, contract DE-AC03-76SF00515}
%\altaffiliation{}
\affiliation{Stanford Linear Accelerator Center,\\
Stanford University, Stanford, CA 94309}

%Collaboration name if desired (requires use of superscriptaddress
%option in \documentclass). \noaffiliation is required (may also be
%used with the \author command).
%\collaboration can be followed by \email, \homepage, \thanks as well.
%\collaboration{}
%\noaffiliation

\date{\today}

\begin{abstract}

We consider the impedance of a structure with rectangular, periodic
corrugations on two opposing sides of a rectangular beam tube.
Using the method of field matching, we find the modes in such a structure.
We then limit ourselves to the
the case of small corrugations, but where the depth of corrugation
is not small compared to the period. For such a structure
we generate analytical approximate
solutions for the wave number $k$, group velocity $v_g$, and loss factor
$\kappa$ for the lowest (the dominant) mode
which, when compared with the results of the complete numerical solution,
agreed well.
We find:
if $w\sim a$, where $w$ is the beam pipe width and
$a$ is the beam pipe half-height, then one mode dominates the impedance, with
$k\sim1/\sqrt{w\delta}$ ($\delta$ is
the depth of corrugation), $(1-v_g/c)\sim\delta$, and $\kappa\sim1/(aw)$,
which (when replacing $w$ by $a$)
is the same scaling as was found for small corrugations in a {\it round} beam
pipe.
Our results disagree in an important way with a recent paper of
Mostacci {\it et al.}
[A. Mostacci {\it et al.}, Phys. Rev. ST-AB, {\bf 5}, 044401 (2002)],
where, for the rectangular structure, the authors
obtained a synchronous mode with the same frequency $k$, but with
$\kappa\sim\delta$.
Finally, we find that if
$w$ is large compared to $a$ then many nearby modes contribute
to the impedance, resulting in a wakefield that Landau damps.

\vskip3.5cm
\centerline {Submitted to Physical Review Special Topics--Accelerators and Beams:}

\centerline {High Brightness 2002 Special Edition}

\end{abstract}

% insert suggested PACS numbers in braces on next line
% insert suggested keywords - APS authors don't need to do this
%\keywords{}

%\maketitle must follow title, authors, abstract, \pacs, and \keywords

\maketitle

\section{INTRODUCTION}

In accelerators with very short bunches, such as is envisioned in
the undulator region of the
Linac Coherent Light Source (LCLS) \cite{LCLS:98}, the
wakefield due to the roughness of the beam-tube walls
can have important implications on the required smoothness and
minimum radius allowed for the beam tube.
One model that has been used to study roughness is a cylindrically-symmetric
structure with small, rectangular, periodic corrugations. For such a structure,
if the depth-to-period ratio of the corrugations is not small compared to 1,
it has been found that the impedance
is dominated by a single strong mode with wave number $k\sim1/\sqrt{a\delta}$,
with $a$ the structure radius and $\delta$ the depth of corrugation,
and loss factor $\kappa=4/a^2$
(in Gaussian units) \cite{Timm:98,BaneNovo:99}.

In a recent report Mostacci {\it et al.} \cite{Mostacci:02}, studied the impedance of a
structure with small, rectangular, periodic corrugations on opposing sides
of a {\it rectangular} beam tube using a perturbation
approach.
For a beam tube with width $w$ comparable to height $2a$ the authors find a mode
with a similar frequency dependence as in the round case,
but with a loss factor that is proportional to the depth of corrugation $\delta$.
If this model is meant to represent surface roughness with {\it e.g.}
$\delta\sim1$~$\mu$m and $a\sim1$~cm, then their result implies a
factor $\sim10^{-4}$ smaller interaction strength than was obtained
in the earlier cylindrically symmetric calculations.
Such a result seems unlikely---we would not expect a huge difference
in loss factor when changing from round to rectangular geometry.
It is the goal of this paper to resolve this discrepancy and to show
that a correct calculation for the rectangular cross section indeed gives a result
that differs only by a numerical factor from the round case.

Another motivation for this work is to
understand the impedance of two corrugated plates, the
limit of our geometry when $w$ becomes large.
And although, when $w$ is not large, the geometry is somewhat artificial,
it may still
be a useful model for some vacuum chamber objects
of accelerators, {\it e.g.}
for the screens in the LHC vacuum chamber \cite{Mostacci:02}.
And thirdly,
we note that fabricating a structure with
artificially large corrugations, for the purpose of experimentally studying roughness
impedance, may be much easier for the rectangular than the round beam pipe.

In this report we calculate the impedance of the rectangular
structure of Mostacci {\it et al.}---but
not limiting ourselves to small corrugations---using
the method of field matching.
The solution is written as an infinite homogeneous matrix equation
that we truncate to solve numerically.
Note that our approach is very similar to that used for the analogous
cylindrically symmetric problem in the computer program TRANSVRS \cite{BaneZot:80}.
Note also that recently, Xiao {\it et al.} used
a similar method to solve the impedance of the rectangular
structure, but with the corrugated surfaces replaced by dielectric
slabs \cite{xiao:01}.
Next, using a perturbation approach applied to the field matching equations
we find the analytical solution for the limit of small corrugations.
Finally, we compare the analytical to the numerical results.

\section{FIELD MATCHING}

\begin{figure}[htbp]
\centering
\includegraphics*[width=115mm]{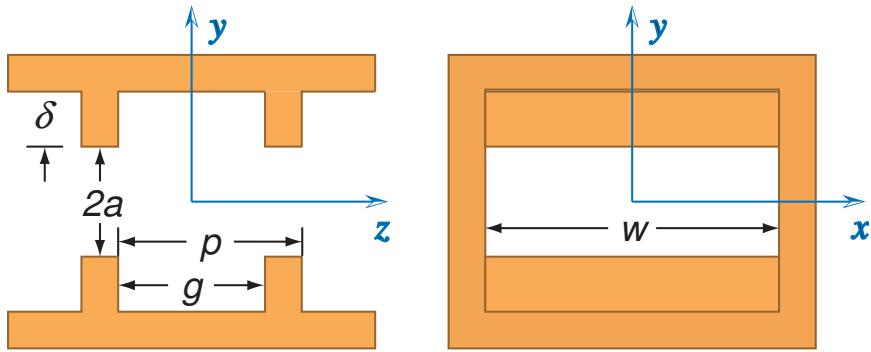}
\caption{
A longitudinal cut of the structure geometry considered here, showing two periods
in the $z$-$y$ plane (left),
and a transverse cut showing the cross-section of the structure (right).
}\label{geom_fi}
\end{figure}

We consider a periodic, rectangular structure with perfectly conducting walls, two
periods of which are sketched in Fig.~\ref{geom_fi}.
In the horizontal ($x$) direction
the structure does not vary, except for walls at
$x=\pm w/2$. One period of the structure extends
longitudinally to $z=\pm p/2$.
This cell can be divided into two regions:
Region~I, the ``tube region'', extends to
$y=\pm a$; Region~II, the ``cavity region'', for $z=\pm g/2$,
extends beyond $y=\pm a$ to $y=\pm (a+\delta)$. An exciting point beam
moves at the speed of light $c$
from minus to plus infinity along the $z$ axis.
We are interested in the steady-state fields excited by the beam, and
assume that initial transients have all died down.
Note that
we will work in Gaussian units throughout.

We assume that the fields of a mode excited by the beam have a
time dependence $e^{jkct}$, where $k$ is the mode wave number and
$t$ is time.
For either region the fields can be obtained from two Hertz
vectors, ${\bf\Pi}_m$ and ${\bf\Pi}_e$, which generate,
respectively, TM and TE components of the fields:
\begin{eqnarray}
{\vec {\cal E}}&=& \nabla\times\nabla\times{\bf\Pi}_m-jk\nabla\times{\bf\Pi}_e\quad,\\ \nonumber
{\vec {\cal H}}&=& \nabla\times\nabla\times{\bf\Pi}_e+jk\nabla\times{\bf\Pi}_m\quad.
\end{eqnarray}
Since there is no variation in the $x$ direction
we choose it as the direction of the Hertz vectors.
To satisfy the boundary conditions at $x=\pm w/2$
the fields vary as cosines and sines of $k_x x$
where
\begin{equation}
k_x={m\pi\over w}\quad,
\end{equation}
 with $m$ an odd integer (see below).
The general solution involves a summation, over all $m$, of such modes.

Consider modes with horizontal mode number $m$.
In the tube region, the most general form of the
($x$ component of the) Hertz vectors,
consistent with the (perfectly conducting) walls at $x=\pm w/2$, and the
Floquet condition in $z$ is:
\begin{eqnarray}
\Pi_{mx}^I&=&\sum_{n=-\infty}^\infty\left[A_n\sinh(k_{yn}^I y)
+B_n\cosh(k_{yn}^I y)\right]
\sin (k_x x)\, e^{-j\beta_n z}\quad,\\ \nonumber
\Pi_{ex}^I&=&\sum_{n=-\infty}^\infty\left[C_n\sinh(k_{yn}^I y)
+D_n\cosh(k_{yn}^I y)\right]
\cos (k_x x)\, e^{-j\beta_n z}\quad,
\label{piI_eq}
\end{eqnarray}
with
\begin{equation}
\beta_n=\beta_0+{2\pi n\over p}\quad,\quad\quad
{k_{yn}^I}=\sqrt{\beta_n^2-k^2+k_x^2}\quad.
\end{equation}
%;at the end we may need to sum over $m$.
Since the structure is symmetric in $y$ about $y=0$, the field components will
be either even or odd in $y$, and the modes will split into two categories.
In the first type $A_n=D_n=0$ and the resulting modes have
${\cal E}_z\neq0$ on axis,
in the second type $B_n=C_n=0$ and the resulting modes have
${\cal E}_z=0$ on axis.
In either case we are left with only 2 sets of
unknown constants in Region~I. Since an on-axis beam can only excite
modes of the first type, it is this type in which we are interested.

In the cavity region, the most general form of the Hertz potentials,
consistent with perfectly conducting boundary conditions at $z=\pm g/2$ and
$y=\pm (a+\delta)$ is:
\begin{eqnarray}
\Pi_{mx}^{II}&=&\sum_{s=1}^\infty E_s\sin[k_{ys}^{II} (a+\delta-y)]
\sin (k_x x) \sin[\alpha_s(z+g/2)]\quad,\\ \nonumber
\Pi_{ex}^{II}&=&j\sum_{s=0}^\infty F_s\cos[k_{ys}^{II} (a+\delta-y)]
\cos (k_x x) \cos[\alpha_s(z+g/2)]\quad,
\end{eqnarray}
with
\begin{equation}
\alpha_s={\pi s\over g}\quad,\quad\quad
{k_{ys}^{II}}=\sqrt{k^2-\alpha_s^2-k_x^2}\quad.
\end{equation}
Note that in both regions ${\cal E}_y$, ${\cal E}_z$, and ${\cal H}_x$ depend
on $x$ as $\cos(k_x x)$ and, therefore, the boundary conditions on the walls at
$x=\pm w/2$ are automatically satisfied.

We need to match the tangential electric and magnetic fields in the
matching planes, at $y=\pm a$:
\begin{eqnarray}
{\cal E}_{z,x}^I&=&  \left\{ \begin{array}{r@{\quad\ :\quad}l}
                {\cal E}_{z,x}^{II}&\quad\quad\ \ \,|z|<g/2\\0&g/2<|z|<p/2\end{array}\right. \\
{\cal H}_{z,x}^I&=& \quad {\cal H}_{z,x}^{II}\ \ :\quad\quad\quad\ \, |z|<g/2\quad.
\end{eqnarray}
Using the orthogonality of $e^{-j\beta_n z}$ over
$[-p/2,p/2]$ in Region~I, and
$\sin[\alpha_s(z+g/2)]$ and $\cos[\alpha_s(z+g/2)]$ over
$[-g/2,g/2]$ in Region~II, we obtain a matrix system that we truncate
to dimension $2(2{\cal N}+1)\times 2(2{\cal N}+1)$, where ${\cal N}$ is the largest value of $n$
that is kept. To obtain modes excited by the beam we need to set
$\beta_n=k$ for one value of $n$.
The frequencies at which the determinant of the resulting matrix
vanishes are the excited frequencies of the structure.

The relation of the
coefficients at the excited frequencies
gives the eigenfunctions of the modes,
from which we can then obtain the $(R/Q)$'s and the loss factors.
The loss factor, the amount of energy lost to a mode per unit charge
per unit length of structure, is given by
\begin{equation}\label{kloss_general}
\kappa={|{\cal E}_{zs}|^2\over 4u(1-v_g/c)}\quad,
\end{equation}
with ${\cal E}_{zs}$ the synchronous component of the longitudinal field
on axis, $u=(8\pi p)^{-1}\int|{\cal E}|^2 dx\,dy\,dz$,
the (per unit length) stored energy in the mode
[the integral is over the volume of one period of structure], and
$v_g$ the group velocity in the mode.
Note that the factor $1/(1-v_g/c)$ is often
neglected in loss factor calculations
(it appears to have been neglected in Mostacci {\it et al.}).
This factor in the loss factor, which---as we will see---is very important
in structures with small corrugations, is discussed in
Refs.~\cite{Chojnacki:93,Millich:99,Wuensch:99}; we give a new
derivation of it in Appendix~A.
Finally the longitudinal wakefield is given as
\begin{equation}\label{long_wake}
W(s)=2\Phi(s)\sum_n\kappa_n\cos(k_ns)\quad,
\end{equation}
with $\Phi(s)=0$ for $s<0$, $1$ for $s>0$, and the sum is over all
excited modes.

In Appendix~B we present more details of the calculation of the
modes of the corrugated
structure using field matching. We have
written a Mathematica program that
numerically solves these equations
for arbitrary corrugation size. The results of this program will be used
to compare with small corrugation approximations presented
in the following section.

\section{Small Corrugations}

Let us consider the case where the corrugations are small, but with
$\delta\sim g\sim p\ll a\sim w$. In the analogous cylindrically symmetric
structure it was found that: (i)~there is one dominant mode (its loss factor
is much larger than those of the other modes),
(ii)~this mode has a low phase advance per cell, and
(iii)~the frequency of the mode $k\sim1/\sqrt{a\delta}$
\cite{BaneNovo:99,BaneStupakov:00}.
For our rectangular structure we look for a mode with the same properties.
As was the case for the cylindrically
symmetric problem we also assume that
the fields in the cavity region are approximately independent
of $z$, and that one term in the expansion of the $\vec\Pi$ vectors,
the term with $n=0$ and $s=0$, suffices to give a consistent solution
to the field matching equations \cite{BaneNovo:99}.
Note that, it is true that to match
the tangential fields well on the matching plane may require many
space harmonics (though even then, near the corners,
Gibbs phenomena and the edge condition will result in poor convergence);
nevertheless, as with the analogous cylindrically symmetric problem,
the global mode parameters in which we are most interested---frequency
$k$, group velocity $v_g$, and loss factor
$\kappa$---can be obtained to good approximation when keeping only the one
(the $n=0$, $s=0$) term.
% term if the matching equations
%give consistent results.

Setting $\alpha=0$ implies that $\Pi_{mx}^{II}=0$, and that there are only
3 non-zero field components in the cavity region: ${\cal E}_z^{II}$, ${\cal E}_y^{II}$, and
${\cal H}_x^{II}$. For small corrugations the excited modes become approximately TM modes.
To allow matching at the interface of Regions~I and II
we end up with
\begin{eqnarray}
\Pi_{mx}^I&\approx&0\quad,\\ \nonumber
\Pi_{ex}^I&\approx& C_0\sinh(k_{y0}^I y)
\cos (k_x x)\, e^{-j\beta_0 z}\quad,
\end{eqnarray}
and
\begin{eqnarray}
\Pi_{mx}^{II}&\approx&0\quad,\\ \nonumber
\Pi_{ex}^{II}&\approx& jF_0\cos[k_{y0}^{II} (a+\delta-y)]
\cos (k_x x)\quad.
\end{eqnarray}

Let us sketch how we match the fields:
We equate ${\cal E}_z$ and ${\cal H}_x$ for the two
regions at $y=\pm a$; we multiply the first equation
by $e^{j\beta_0z}$ and integrate over one period in $z$,
and then we integrate the second equation
over the gap in $z$.
When we divide the resulting equations one by the other, the constants $C_0$, $F_0$,
drop out, and we are left with an approximation to
the dispersion relation, one valid in the
vicinity of the synchronous point
(the subscript 0 for $\beta$ is understood):
\begin{equation}
\sqrt{\beta^2-k^2+k_x^2}\,\coth\left(\sqrt{\beta^2-k^2+k_x^2}\, b\right)=
{4\sin^2({\beta g/2})\over gp\beta^2}\sqrt{k^2-k_x^2}
\tan\left(\sqrt{k^2-k_x^2}\,\delta\right)
\quad.\label{disp_eq}
\end{equation}

To properly keep track of the
relative size of the terms in further calculations,
we assign to each parameter an order
using the small parameter $\epsilon$: let
$a$, $w$, be of order 1; $\delta$, $g$, $p$, of order $\epsilon^2$;
and $k$, $\beta$, of order $1/\epsilon$.
To find the synchronous frequency we let $\beta=k$ in
Eq.~(\ref{disp_eq}), expand the equation
to lowest order in $\epsilon$, and then set $\epsilon=1$.
The result is
\begin{equation}
k^2_m={k_x p\over \delta \,g}
\coth\left({k_x a}\right)\quad,\label{ksync_eq}
\end{equation}
(the subscript $m$ is included here to remind us of the $m$ dependence).
Note that, if $a\sim w$ (and $p\sim g$)
then $k\sim1/\sqrt{w\delta}$, which is of the
same order as the result that was found for the cylindrically symmetric problem.
Note also that, for the limit
$g=p$, the dispersion relation
and the synchronous frequency here agree with those given in Mostacci {\it et al.}

\begin{figure}[htbp]
\centering
\includegraphics*[width=95mm]{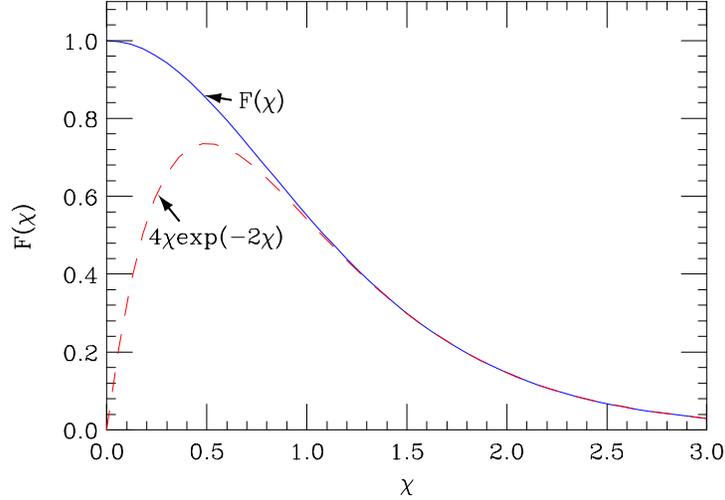}
\caption{
The function $F(\chi)$ (solid) and the approximation
$4\chi e^{-2\chi}$, valid for $\chi \gtrsim 1$ (dashes).
}\label{f_fi}
\end{figure}

For the group velocity we take the partial derivative
of Eq.~(\ref{disp_eq}) with respect to $\beta$, and rearrange terms
to obtain $1-\partial k/\partial\beta=1-v_g/c$. After expanding
in $\epsilon$, keeping the lowest order term, and finally setting $\epsilon=1$
we obtain
\begin{equation}
\left(1-{(v_g)_m\over c}\right)={2\delta\, k_x g\over p}\left[
{\sinh^2(k_x a)\over\sinh(k_x a)\cosh(k_x a)-k_x a}\right]\quad.\label{vg_eq}
%\coth(k_x a)-k_x a\,{\rm csch}^2(k_x a)\right]^{-1}
\end{equation}
Note that, as in the cylindrically symmetric problem,
 $(1-v_g/c)\sim\delta$.
The loss factor of our structure
\begin{equation}
\kappa_m=
{2\pi\over wa}F\left({k_x a}\right)\quad,
\label{kloss_eq}
\end{equation}
with
\begin{equation}
%F(\chi)=\chi{\rm csch}(\chi){\rm sech}(\chi)\quad.
F(\chi)={\chi\over\sinh(\chi)\cosh(\chi)}\quad.
\end{equation}
The function $F(\chi)$ and an approximation for large $\chi$ are
shown in Fig.~\ref{f_fi}.
Note that for $\kappa$ in the MKS units of [V/pC/m],
one multiplies Eq.~(\ref{kloss_eq})
by the quantity $Z_0c/(4\pi)$,
with $Z_0= 377$~$\Omega$. Note also that our result is independent
of $\delta$, unlike the result of Mostacci {\it et al.}

The total longitudinal wakefield
is given by
Eq. (\ref{long_wake}).
Note that, if $w\lesssim a$ ($\chi\gtrsim1$) then one mode dominates the wake,
just like in the round case.
(For example, if $\chi=1$, then the amplitude of the first, $m=1$ term
is 20 times larger than that of the next, $m=3$ term in the wake sum.)
If, however, $w\gg a$,
then more than one mode will contribute to the impedance of the structure;
in the limit of $w\rightarrow\infty$ (two corrugated plates)
there will be a continuum of modes
contributing to the impedance.
The impedance is given by the Fourier transform of the wake.
Its real part is
\begin{equation}
{\cal R}{\rm e}Z=\pi\sum_m \kappa_m\left[\delta(\omega-k_m c)
+\delta(\omega+k_m c)\right]\quad.\label{rez_eq}
\end{equation}
%Note that the surface impedance, the ratio of ${\cal E}_z/{\cal H}_x$ at $y=a$,
%is given by $jk_x k\coth(k_x a)/(k^2-k_x^2)$.

\begin{figure}[htbp]
\centering
\includegraphics*[width=157mm]{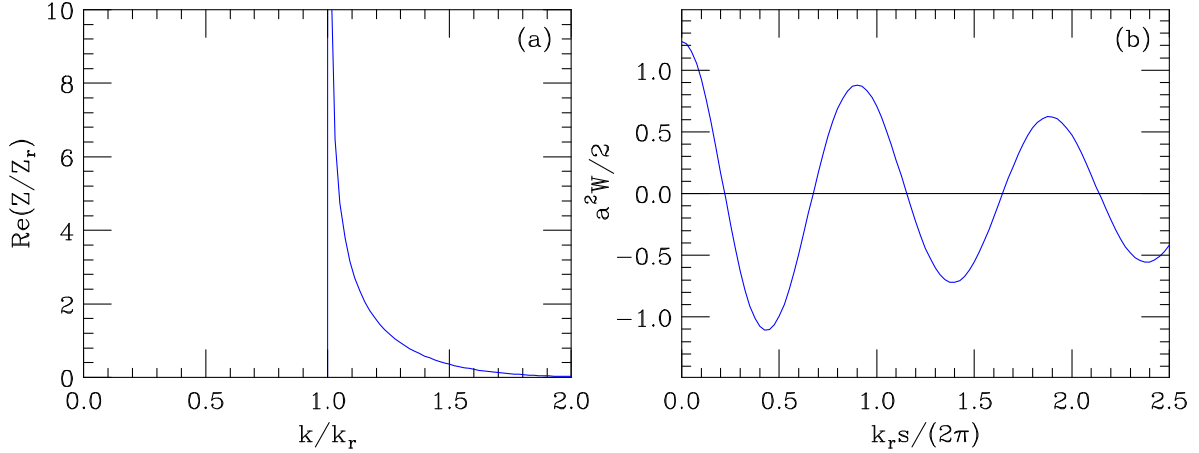}
\caption{
For the case of two corrugated plates ($w\rightarrow\infty$):
${\cal R}{\rm e}(Z)$~(a) and the wake~(b), with
$k_r=\sqrt{p/(a\delta g)}$ and $Z_r=\pi/(a^2 k_r c)$.
}\label{imp_fi}
\end{figure}

Consider now the limit of two corrugated plates ($w\rightarrow\infty$).
The mode spectrum becomes continuous and the sum
in Eq.~(\ref{rez_eq}) can be replaced by an integral
\begin{equation}
{\cal R}{\rm e}Z={\pi\over a^2}\int_0^\infty d\chi\,F(\chi)
\left[\delta\left(\omega- c\sqrt{{ p\over a\delta g}\chi\coth(\chi)}\right)
+\delta\left(\omega+ c\sqrt{{ p\over a\delta g}\chi\coth(\chi)}\right)\right]\quad.
\end{equation}
The integral can be solved numerically, with the use of the relation
$\int dx\,g(x)\delta[f(x)]=[g(x)/|f^\prime(x)|]_{x=x0}$ where $f(x_0)=0$.
The result is shown in Fig.~\ref{imp_fi}a; note that
the axes are normalized to $k_r=\sqrt{p/(a\delta g)}$
and $Z_r=\pi/(a^2 k_r c)$.
We see a continuous spectrum of modes beginning
at wave number $k_r$, with average $1.14k_r$ and
rms $0.18k_r$.
The corresponding wakefield becomes a damped oscillation
(see Fig.~\ref{imp_fi}b). We see an
effective $Q\sim10$. Note that $W(0^+)=\pi^2/(4a^2)$
[to be discussed more in a later section].

Finally, we should point out that
it has been observed for the case of the cylindrically symmetric
problem that, if the small corrugations are replaced by a
thin dielectric layer of thickness $\delta$, and if
the correspondence is made that the dielectric
constant $\epsilon=p/(p-g)$, then the results
for the two problems are the same \cite{BaneNovo:99}.
Recently the modes in a rectangular structure of Fig.~\ref{geom_fi},
but with the corrugated surfaces replaced by
dielectric slabs, have been obtained by Xiao {\it et al.},
also using a field matching approach \cite{xiao:01}. If we take their
results, letting the thickness of the dielectric layers ($\delta$) be small,
we obtain our results for $k$, $v_g$, and $\kappa$ when we make
the correspondence $\epsilon=p/(p-g)$.

\subsection{Comparison with Numerical Results}

To test the validity of the analytical approximations
in the case of small corrugations, we compare with numerical
results obtained by the
Mathematica field matching program (the method of solution
is described in Appendix~B).
Consider as an example a square
beam tube ($w/a=2$) with
$p/a=0.05$, $g/a=0.025$, and $\delta/a=0.025$, and
let us consider the lowest ($m=1$) mode. In the
field matching program we take ${\cal S}=4$ and ${\cal N}=4$,
{\it i.e.} 5 space harmonics are kept in the cavity region and
9 in the tube region.
(We find that, for the example geometry,
keeping more terms has no significant effect on the results.)

\begin{figure}[htbp]
\centering
\includegraphics*[width=95mm]{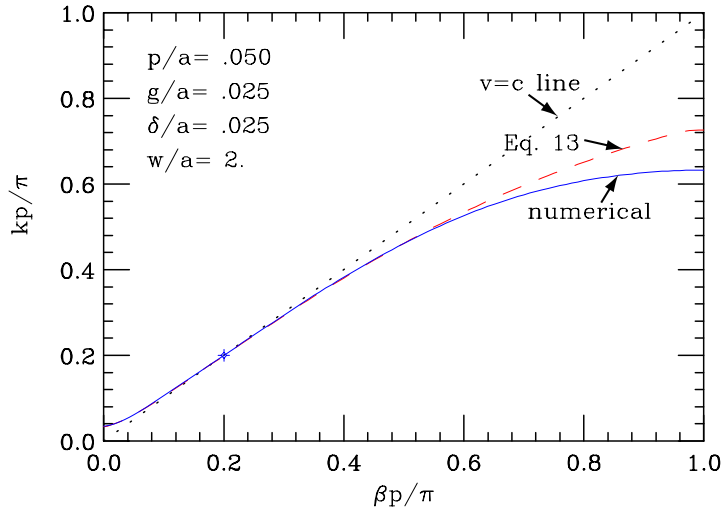}
\caption{
A dispersion curve example: shown are the numerical result (solid),
the synchronous point (the cross plotting symbol),
and the approximation, Eq.~(\ref{disp_eq}) (dashes).
Also shown is the speed of light line (dots).
}\label{disp_fi}
\end{figure}

We begin by comparing the dispersion curve (see Fig.~\ref{disp_fi}).
Shown are the field matching result (the solid curve) and
the approximation, Eq.~(\ref{disp_eq}) (the dashes). We see that the two
agree well except far from the synchronous phase.
The cross plotting symbol locates the
synchronous point, with $k p=0.200\pi$, a result
which is 7.5\% larger than the analytical value
of Eq.~(\ref{ksync_eq}).
It is interesting to note that this dispersion curve is almost identical to the one
obtained (also by field matching)
for the same geometry but in a {\it round} beam pipe \cite{BaneNovo:99}.
As for the loss factor, we find that it is a factor 0.84 as large as the
analytical approximation, Eq.~\ref{kloss_eq}.

These results confirm the validity of the analytical approximations
for the structure with small corrugations, provided that
the depth of corrugation $\delta$ is
not small compared to the corrugation period $p$.
However, in Ref.~\cite{BaneNovo:99} it was shown that for the analogous round
structure the corresponding analytical formulas break down when
$\delta$ becomes small compared to $p$: as $\delta$ decreases
the frequency first increases than decreases as compared to the analytical
result; meanwhile the loss factor continually decreases.
When $\delta$ is small compared to $p$
the impedance is no longer well characterized by a single resonance,
and is best described by a different model \cite{stupakov}.
As expected,
we find the same kind of behavior in our rectangular structure.
If, for example, we reduce $\delta$ in our
example problem by a factor of 2, we find that the frequency becomes
18\% larger, and the loss factor 30\% smaller, than the
values given by the analytical formulas.

\subsection{Discussion}

Our result for the loss factor, Eq.~(\ref{kloss_eq}), is independent of the
depth of corrugation $\delta$, as was found previously
for the analogous cylindrically symmetric
problem \cite{BaneNovo:99,BaneStupakov:00}.
This result, however, is in disagreement
with the result of Mostacci {\it et al.}, where the loss factor was
found to be directly proportional to $\delta$.
%The discrepancy appears to come from the fact that
%the $1/(1-v_g/c)$ factor was neglected by Mostacci {\it et al.}
This discrepancy is important to resolve.

%What confidence do we have that our answer is correct?
%We have additional evidence that our solution is correct:%
There is a general relation that holds for the wake
directly behind the driving particle
\begin{equation}
W(0^+)={2\over \pi}
\int_{0}^\infty{\cal R}{\rm e} Z(\omega)\,d\omega=2\sum_m \kappa_m\quad,
\end{equation}
a relation that does not depend on the specific boundary conditions at the wall.
To discuss it, consider first the analogous cylindrically symmetric
problem.
It was earlier found that, as long as the corrugations are small
and the depth $\delta\gtrsim p$, the contribution of one mode
dominates the wake sum.  In this case,
it was found that, as here, $W(0^+)$ (or $\kappa$)
is independent of $\delta$ \cite{BaneNovo:99}.
If the corrugations are replaced by a thin dielectric layer,
$W(0^+)$ does not depend on the dielectric properties
(neither $\delta$ nor $\epsilon$) \cite{Novokhatski:97}.
In the same way, if the corrugations
are replaced by
a lossy metal, $W(0^+)$ will not depend on the conductivity \cite{Chao:93}.
And in all three cases the answer is the same: $W(0^+)=4/a^2$.
[In fact, this relation is also valid for the (steady-state) wake
of a periodic accelerator structure,
with $a$ the iris radius \cite{Gluckstern:89,comment1}.]

\begin{figure}[htbp]
\centering
\includegraphics*[width=95mm]{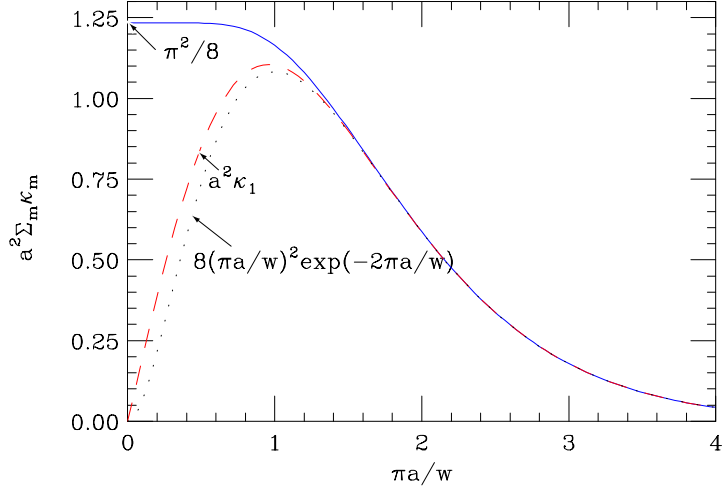}
\caption{
The sum of the loss factors
$a^2\sum_m \kappa_m$ [$=a^2W(0^+)/2$] as function
of $\pi a/w$ (solid). Also shown are the contribution
of the first mode, $a^2\kappa_1$ (dashes), and
the approximation $8(\pi a/w)^2\exp(-2\pi a/w)$ (dots).
}\label{kloss_fi}
\end{figure}

We expect the same type of behavior to hold in a corrugated, rectangular structure,
{\it i.e.} that $W(0^+)$ depends only on the cross-section geometry of
the beam pipe.
In Fig.~\ref{kloss_fi} we plot, for our rectangular structure,
$a^2W(0^+)/2=a^2\sum_m\kappa_m$, as function of $\pi a/w$
(the solid curve).
Also shown
is the contribution of only the first ($m=1$) term (dashes),
and the approximation $8(\pi a/w)^2\exp(-2\pi a/w)$ (dots).
Note that, for $\pi a/w$ small, many modes contribute
to the sum; for $\pi a/w\gtrsim1$, one
mode dominates.
As with the cylindrically-symmetric case,
$W(0^+)$ must still be correct if we replace the corrugated surfaces
by thin dielectric slabs, or by lossy metal plates.
We know of no published result for $W(0^+)$ in our rectangular geometry
to compare with;
nevertheless, Henke and Napoly found $W(0^+)$ between two resistive parallel
plates \cite{Henke:90}, which becomes
the limit of our geometry as $w\rightarrow\infty$.
Their result, $a^2W(0^+)/2=\pi^2/8$, agrees with our
calculation for $\pi a/w\rightarrow0$, and confirms our result.

For a given bunch shape and fixed $\delta/p$,
as the depth of corrugation $\delta$ decreases, we expect the induced
voltage (the convolution of the bunch shape with the wake) to also decrease.
If the loss factor does not depend on $\delta$ how does this happen?
The answer is that as $\delta$ decreases, the mode frequency $k$ increases,
and the wake, when convolved with the bunch shape, will yield an
induced voltage that will decrease (at least as fast as $1/k$).
Concerning this question, the
wake of this structure behaves similarly to
the resistive wall wake for very short bunches as the conductivity
increases:
$W(0^+)$ also does not change but the wake first
zero crossing moves closer to $s=0$.

\section{CONCLUSION}

We studied the impedance of a structure with rectangular, periodic
corrugations on two opposing sides of a rectangular beam tube using the
method of field matching. We described a formalism
that, for arbitrary corrugation size, can find the resonant
frequencies $k$, group velocities $v_g$, and loss factors $\kappa$. In addition, for
the case of small corrugations, but where the depth of corrugation
is not small compared to the period, we generated analytical perturbation
solutions for $k$, $v_g$, and $\kappa$ for the dominant mode.
We then compared,
for such a structure,
the results of the computer program and the analytical formulas, and found
good agreement.

In general,
we found that, for the structure of interest, the results are very similar
to what was found earlier for a structure consisting of small corrugations
on a {\it round} beam pipe:
if $w\sim a$, where $w$ is the beam pipe width and
$a$ is the beam pipe half-height, then one mode dominates the impedance, with
$k\sim1/\sqrt{a\delta}$ ($\delta$ is
the depth of corrugation), $(1-v_g/c)\sim\delta$, and $\kappa\sim1/a^2$.
If, however, $w$ is large compared to $a$ we find that many nearby modes contribute
to the impedance, resulting in a wakefield that Landau damps.

\appendix
\section{Excitation of a synchronous mode by a moving relativistic point
charge}

Consider first a cavity of frequency $\omega$ with the electric
field of an eigenmode
    $
    \vec{\cal {E}}(\vec{r})
    e^{j\omega t}.
    $
The energy in the eigenmode is denoted by $U$. If a point charge
$q$ passes through the cavity, it excites this mode to the
amplitude $A_s$ (where $A_s$ is a complex number), so that after
the passage through the cavity the electric field of the mode will
be
    $
    A_s\vec{\cal {E}}(\vec{r})
    e^{j\omega t},
    $
and the energy lost by the charge is equal to $|A_s|^2U$. In
quantum language, this is \emph{spontaneous} radiation of the
charge into the mode under consideration which is indicated by the
subscript $s$. It is clear that $A_s$ is proportional to the
charge of the particle $q$.

To calculate the amplitude $A_s$, let us consider a situation
when, before the charge enters the cavity, the latter already has
this mode excited by an external agent (RF source) to the
amplitude $A_0$. Due to linearity of Maxwell's equation, after the
passage of the charge, the field in the cavity will be equal to
the sum of the initial mode $A_0$ and the spontaneously radiated
mode $A_s$, with the energy given by $|A_s+A_0|^2U$. The change of
the energy $\Delta W$ in the cavity is
    \be\label{energy}
    \Delta W
    =
    |A_s+A_0|^2U
    -
    |A_0|^2U
    =
    (A_s A_0+\mathrm{c.c.})U
    +
    |A_s|^2U
    \,,
    \ee
where c.c. denoted a complex conjugate. Let us consider the limit
of small charges, $q\rightarrow 0$, then we can neglect the last
term on the right hand side of Eq. (\ref{energy}), which scales as
$q^2$, and keep only the first term that is linear in $q$,
    \be\label{energy1}
    \Delta W
    =
    (A_s A_0+\mathrm{c.c.})U
    \,.
    \ee
Discarding the term $\propto q^2$ means that we neglect the
\emph{beam loading} effect.

We can now balance the energy change $\Delta W$ with the work done
by the external field $A_0$ during the passage of the charge. This
work is equalt to the integral of the electric field ${\cal
{E}}_z(z)$ along the particle's orbit
    \be\label{work}
    \Delta W
    =
    -q
    {\cal R}{\rm e}
    A_0
    \int
    dz {\cal {E}}_z(z)
    e^{j\omega z/v}
    \nonumber =
    -
    \frac{q A_0}{2}
    \int
    dz {\cal {E}}_z(z)
    e^{j\omega z/v}
    +\mathrm{c.c.}
   \ee
Comparing Eq. (\ref{energy1}) with Eq. (\ref{work}) we conclude
that
    \be\label{A}
    A_s
    =
    -
    \frac{q}{2U}
    \int
    dz {\cal {E}}_z(z)
    e^{j\omega z/v}.
   \ee
Hence we found the amplitude of spontaneous radiation of the
particle in terms of the integral along the particle's orbit of
the electric field.

The energy lost by the particle (loss factor) is
    \be
    |A_s|^2U=
    \frac{q^2 |V|^2}{4U},
    \ee
where the \emph{voltage} $V=\int dz {\cal {E}}_z(z) e^{j\omega
z/v}$.

Let us now apply the same approach as above to the excitation of a
mode that propagates with the speed of light in a waveguide. To
deal with a mode of finite energy we consider a wave packet, and
assume that the packet has a length $L$, as shown in Fig.
\ref{fig:wave_packet} below.
    \begin{figure}[htb]
    \centering\scalebox{.8}{
    \includegraphics*[width=80mm]{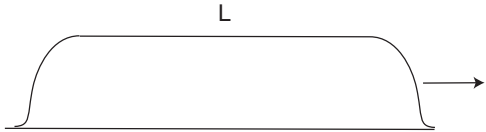}}
    \caption{The shape of the wave packet of the synchronous mode.
    The packet has a long plateau of length $L$ and short edges.
    \label{fig:wave_packet}}
    \end{figure}
It propagates in the pipe with the group velocity $v_g$. The
energy in the mode $U$ can be related to the energy flow $P$
(integrated over the cross section averaged over time the Pointing
vector) if we note that $U/L$ is the energy per unit length, and
hence $(U/L)v_g$ is the energy flow equal to $P$, hence
    \be
    U
    =
    \frac{PL}{v_g}
    \,.
    \ee

Now, the particle is synchronous with the wave and stays all the
time in the same phase, so it sees the same longitudinal electric
field ${\cal {E}}_z$ which we denote by ${\cal E}_{zs}$. The integral
from Eq. (\ref{A}) can be written as
    \be
    \int
    dz {\cal {E}}_z(z)
    e^{j\omega z/v}
    \rightarrow
    cT {\cal E}_{zs}
    \,,
    \ee
where $T$ is the interaction time between the wave and the
particle. This is actually the time when the particle stays in the
wave, and taking into account that the wave is moving with
velocity $v_g$ and the particle is moving with $c$
    \be
    T
    =
    \frac{L}{c-v_g}.
    \ee
Hence, for the amplitude of the radiated wave we find
    \be\label{A_sync_mode}
    A_s
    =
    -
    \frac{q}{2U}
    {\cal E}_{zs}
    \frac{cL}{c-v_g}
    \,,
   \ee
and the energy $W$ radiated by the particle
    \be
    W
    =
    A_s^2 U
    =
    \frac{q^2}{4U}
    {\cal E}_{zs}^2
    \frac{c^2L^2}{(c-v_g)^2}.
    \ee

To find the energy radiated per unit length of the path, we divide
$W$ by the length of the interaction path $Lc/(c-v_g)$, which
gives
    \be
    \frac{dW}{dz}
    =
    \frac{q^2}{4u}
    {\cal E}_{zs}^2
    \frac{c}{(c-v_g)},
    \ee
where the energy per unit length of the path $u=U/L$. Finally,
since the loss factor $\kappa = q^{-2}dW/dz$, we arrive
at Eq. (\ref{kloss_general}).

\section{FIELD MATCHING, THE GENERAL SOLUTION}

In Section~II we presented Hertz vectors and wave numbers for Regions~I
and II, and also the four equations that need to be matched at the interface
$y=\pm a$.
We continue with the notation introduced there:
We multiply the matching equations for ${\cal E}_z$ and ${\cal E}_x$
by $e^{j\beta_{n^\prime} z}$ and integrate over
$[-p/2,p/2]$; and
we multiply the matching equations for ${\cal H}_z$ and ${\cal H}_x$ by
$\sin[\alpha_{s^\prime}(z+g/2)]$ and $\cos[\alpha_{s^\prime}(z+g/2)]$
and integrate over
$[-g/2,g/2]$.
We obtain the infinite set of equations:
\begin{eqnarray}
(C^\prime_nk k_{yn}^I w-B^\prime_nm\pi\beta_n)\cosh(k_{yn}^I b)&=&
{g\over p}
\sum_sN_{ns}(-F^\prime_s k k_{ys}^{II} w+E^\prime_s m\pi\alpha_s)\sin(k_{ys}^{II}\delta)\nonumber\\
-B^\prime_n\cosh(k_{yn}^I b)&=&{g\over p}
\sum_sM_{ns}E^\prime_s\sin(k_{ys}^{II}\delta)\nonumber\\
(E^\prime_s k k_{ys}^{II} w+F^\prime_s m\pi\alpha_s)\cos(k_{ys}^{II}\delta)&=&
2\sum_nM_{sn}(B^\prime_nk k_{yn}^I w-C_nm\pi\beta_n)\sinh(k_{yn}^I b)\nonumber\\
(1+\delta_{s0})F^\prime_s\cos(k_{ys}^{II}\delta)&=&
-2\sum_nC^\prime_nN_{sn}\sinh(k_{yn}^I b)\quad.
\end{eqnarray}
Here
\begin{equation}
B^\prime_n, C^\prime_n=jB_n, jC_n\quad,\quad\quad\quad
E^\prime_s,F^\prime_s=  \left\{ \begin{array}{r@{\quad:\quad}l}
                E_s,F_s&s\ {\rm even}\\
                jE_s,jF_s&s\ {\rm odd}\end{array}\right. \ ,
\end{equation}
\begin{equation}
\left\{\begin{array}{c}
N_{ns}\\ M_{ns}\end{array}
\right\}= \left\{\begin{array}{c}
\beta_n\\ \alpha_s\end{array}
\right\}{2\over(\beta_n^2-\alpha_s^2)g}
  \left[ \begin{array}
          {r@{\quad:\quad}l}
   \sin (\beta_n g/2) & s\ {\rm even} \\
   \cos (\beta_n g/2) & s\ {\rm odd}\end{array}\right.\ ,
\end{equation}
and $\delta_{ss^\prime}$ the Kronecker delta.

This system of equations can be written as a homogenous matrix equation:
\begin{equation}
\left[\left(\begin{array}{cc}
G(H^2-I^2) & -GH\\ -GH & G\end{array}\right)
+\left(\begin{array}{cc}
N & 0\\ 0 & M\end{array}\right)
\left(\begin{array}{cc}
P(Q^2+R^2)/S & -PQ\\ -PQ/S & P\end{array}\right)
\left(\begin{array}{cc}
N^T & 0\\ 0 & M^T\end{array}\right)\right]\left(
\begin{array}{c}
B^{\prime\prime}\\ C^{\prime\prime}\end{array}\right)=0
\end{equation}
with superscript $T$ indicating the transpose of a matrix.
The diagonal elements of diagonal matrices are:
$G_n=\coth(k_{yn}^I b)/(k k_{yn}^I w)$,
$H_n=m\pi\beta_n$, $I_n=k k_{yn}^I w$;
$P_s=2g\tan(k_{ys}^{II}\delta)/(pk k_{ys}^{II} w)$,
$Q_s=m\pi\alpha_s$, $R_s=k k_{ys}^{II} w$, $S_s=(1+\delta_{s0})$.
Note that the system matrix is real.
The expansion coefficients are:
$B_n^{\prime\prime}=-\sinh(k_{yn}^I b)C^\prime_n$ and
$C_n^{\prime\prime}=\sinh(k_{yn}^I b)(k k_{yn}^I w B^\prime_n-m\pi\beta_n C^\prime_n)$.

To solve the matrix equation we truncate
to dimension $2(2{\cal N}+1)\times 2(2{\cal N}+1)$,
where ${\cal N}$ is the largest value of $n$
that is kept.
Therefore, subscript $n$, representing space harmonic number in
the tube region, runs from $-{\cal N}$ to ${\cal N}$; subscript $s$,
representing space harmonic number in
the cavity region, runs from $0$ to ${\cal S}$, the largest value kept.
Note that the values ${\cal N}$, ${\cal S}$, should be chosen so that
$(2{\cal N}+1)/p\approx({\cal S}+1)/g$. The system matrix $U$ is a function
of $\beta_0$ and of $k$. To find synchronous modes, we need to first
set, for one space harmonic $n^\prime$, $\beta_{n^\prime}=k$ and then
numerically search for the value of $k$ for which the
determinant of $U$ becomes zero. The value $n^\prime$ should be taken
to be the nearest integer to $kp/(2\pi)$. To find values of the dispersion curve,
we, for various values of
$\beta_{n^\prime}$ [where again  $n^\prime$ is the nearest integer to $kp/(2\pi)$],
 numerically search for the value of $k$ for which the
determinant of $U$ becomes zero.

Once we have found the frequency we can find the eigenfunctions, from
which we obtain $|{\cal E}_{zs}|^2$ on axis,
\begin{equation}
|{\cal E}_{zs}|^2=k^2\left|B_{ns}^\prime k_x-C_{ns}^\prime k_y^I\right|^2\quad,
\end{equation}
(where $ns$ represents the synchronous space harmonic) and the energy per unit length $u$.
For example, the stored energy in Region~I is given by
%The stored energy in the two regions are
\begin{eqnarray}
&u^I={\displaystyle{1\over 32\pi p}}{\displaystyle\sum_n} \left(
B_n^{\prime2}\left[2k^2k^2_-a+\left(k^4_-+k_x^2k_y^{I2}+k_x^2\beta_n^2\right)
\sinh(2k_{yn}^I a)/k_{yn}^I\right]\right.\quad\quad\quad\quad
\quad\quad\quad\quad\quad\quad\nonumber\\
&\left.+C_n^{\prime2}\left[-2k^2k^2_-a+\left(k^2k_y^{I2}+k^2\beta_n^2\right)
\sinh(2k_{yn}^I a)/k_{yn}^I\right]
-4B_n^\prime C_n^\prime k k_x\beta_n\sinh(2k_{yn}^I a)
\right)
\end{eqnarray}
with $k_-^2=k^2-k_x^2$,
with a corresponding equation giving the energy stored in Region~II.
%, and
%\begin{eqnarray}
%&u^{II}={\displaystyle{1\over 64\pi p}}{\displaystyle\sum_s} \left(
%E_s^{\prime2}\left[4k^2k^2_-a-2\left(k^4-3k^2k_x^{2}+2k_x^4+2k_x^2\alpha_s^2\right)
%\sin[2k_{ys}^{II} a]\cos[2k_{ys}^{II} (a+\delta)]/k_{ys}^{II}\right]\right.\quad\quad\quad\quad
%\quad\quad\quad\quad\quad\quad\nonumber\\
%&+F_s^{\prime2}\left[4k^2k^2_-a-2k^2\left(k_-^2-2\alpha_s^2\right)
%\sin[2k_{ys}^{II} a]\cos[2k_{ys}^{II} (a+\delta)]/k_{ys}^{II}\right]\nonumber\\
%&\left.+8E_s^\prime F_s^\prime k k_x\alpha_s\sin[2k_{ys}^{II} a]\cos[2k_{ys}^{II} (a+\delta)]
%\right)\quad.
%\end{eqnarray}
Note that for small corrugations, $u^{II}\ll u^I$.
The quantity $1/(1-v_g/c)$ is obtained by first calculating the dispersion
curve, and then finding the slope at the synchronous point numerically.
Knowing $|{\cal E}_{zs}|^2$, $u$, and $1/(1-v_g/c)$ we
can finally obtain the loss factor $\kappa$.

\begin{acknowledgments}
This work was supported by the Department
of Energy, contract DE-AC03-76SF00515.
\end{acknowledgments}

\bibliography{kbane}

\end{document}